\documentclass[letterpaper]{article} 
\usepackage{aaai25}  
\usepackage{times}  
\usepackage{helvet}  
\usepackage{courier}  
\usepackage[hyphens]{url}  
\usepackage{graphicx} 
\usepackage{natbib}  
\usepackage{caption} 
\usepackage{algorithm}
\usepackage{algorithmic}

\usepackage{newfloat}
\usepackage{listings}
\DeclareCaptionStyle{ruled}{labelfont=normalfont,labelsep=colon,strut=off} 
\floatstyle{ruled}
\newfloat{listing}{tb}{lst}{}
\floatname{listing}{Listing}
\usepackage{
    amsmath, amssymb, booktabs, amsfonts, nicefrac, microtype, bbding, multirow, paralist, tabularx, subcaption, pifont, enumitem, newtxtext, xspace, appendix, array, etoolbox, adjustbox, tikz
}

\title{Zero-shot Video Moment Retrieval via Off-the-shelf \\ Multimodal Large Language Models}

\author {
    Yifang Xu\textsuperscript{\rm 1}, 
    Yunzhuo Sun\textsuperscript{\rm 2}, 
    Benxiang Zhai\textsuperscript{\rm 1}, 
    Ming Li\textsuperscript{\rm 3}, 
    Wenxin Liang\textsuperscript{\rm 2}, 
    Yang Li\textsuperscript{\rm 1}, 
    Sidan Du\textsuperscript{\rm 1}
}
\affiliations {
    \textsuperscript{\rm 1}Nanjing University\\
    \textsuperscript{\rm 2}Dalian University of Technology\\
    \textsuperscript{\rm 3}Nanjing University of Information Science and Technology\\
    \{xyf, zbx\}@smail.nju.edu.cn, 
    sunyunzhuo@mail.dlut.edu.cn,wxliang@dlut.edu.cn,\\
    mingli@nuist.edu.cn,
    \{yogo, coff128\}@nju.edu.cn
}

\begin{document}

\maketitle

\begin{abstract}

The target of video moment retrieval (VMR) is predicting temporal spans within a video that semantically match a given linguistic query. Existing VMR methods based on multimodal large language models (MLLMs) overly rely on expensive high-quality datasets and time-consuming fine-tuning. Although some recent studies introduce a zero-shot setting to avoid fine-tuning, they overlook inherent language bias in the query, leading to erroneous localization. To tackle the aforementioned challenges, this paper proposes \textbf{Moment-GPT}, a tuning-free pipeline for zero-shot VMR utilizing frozen MLLMs. Specifically, we first employ LLaMA-3 to correct and rephrase the query to mitigate language bias. Subsequently, we design a span generator combined with MiniGPT-v2 to produce candidate spans adaptively. Finally, to leverage the video comprehension capabilities of MLLMs, we apply Video-ChatGPT and span scorer to select the most appropriate spans. Our proposed method substantially outperforms the state-of-the-art MLLM-based and zero-shot models on several public datasets, including QVHighlights, ActivityNet-Captions, and Charades-STA. 
%
\end{abstract}


\section{Introduction}
\label{sec:intro}

\begin{figure}[t!]
  \centering
  \includegraphics[width=\linewidth]{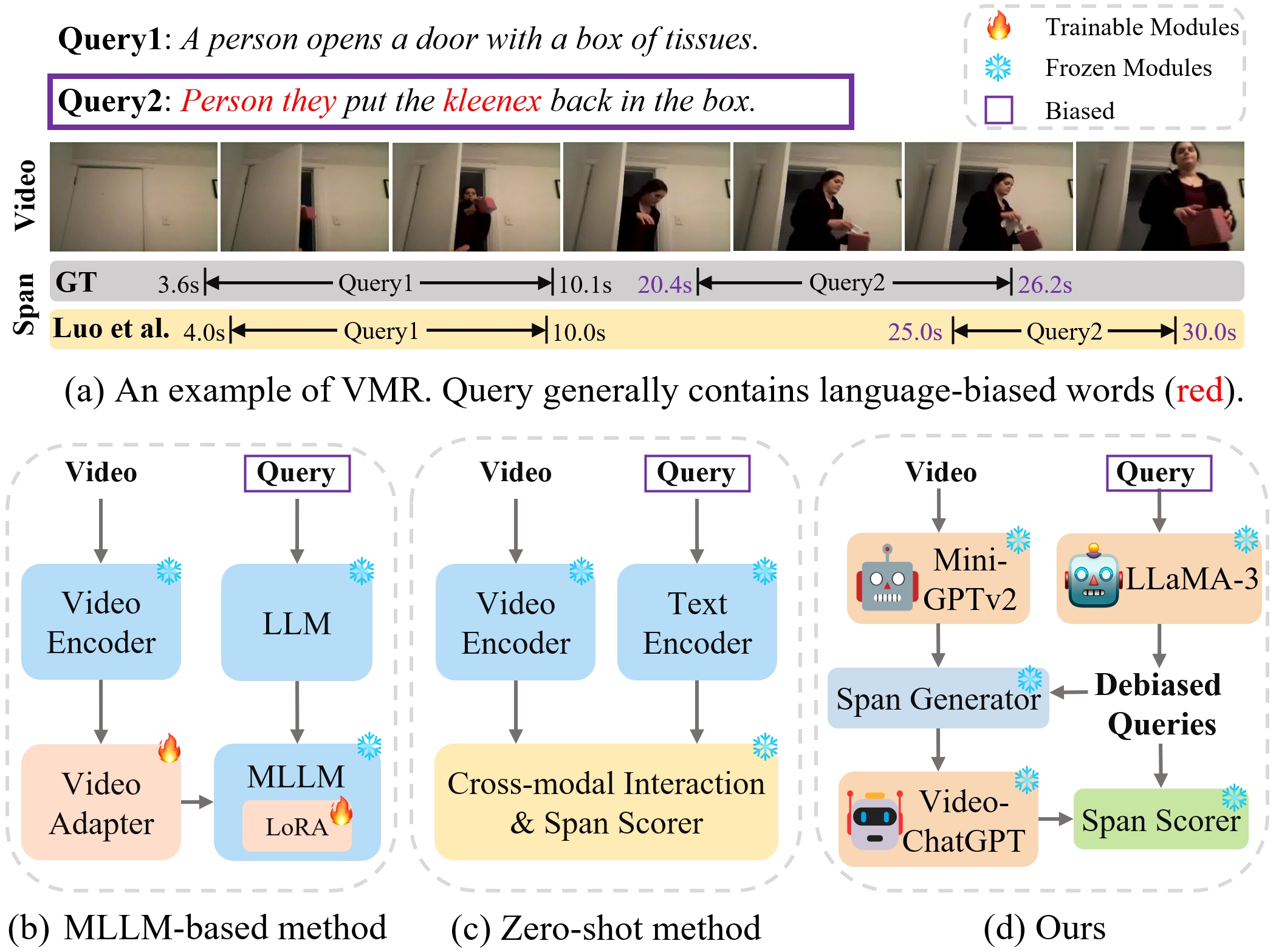}
  \caption{\small
    (a) Illustration of video moment retrieval (VMR). The query containing language bias results in erroneous localization. (b) MLLM-based method demand complex multi-stage fine-tuning using vast volumes of annotated data. (c) Zero-shot method \cite{FVLM-2023} cannot avoid performance degradation caused by language bias. (d) Our proposed Moment-GPT harnesses the video comprehension capabilities of MLLMs 
    without further fine-tuning. It also utilizes LLM 
    to reduce bias, enhancing overall accuracy.
  }
  \label{fig:high-level}
\end{figure}

Video moment retrieval (VMR) is a crucial task in the field of video understanding, attracting widespread attention in the last few years owing to its potential applications in video surveillance and human-computer interaction \cite{lyuNovelTemporalMoment2023, le2010surveillance, yan2024mlp, PFANet-2021, DAFFNet-2021, Hifi-portrait, HP3-2025}. It aims to locate temporal spans (segments) that are semantically related to a specified sentence query from an untrimmed video, with each span comprising a beginning and an ending moment. Fig.~\ref{fig:high-level}~(a) presents an instance of VMR.

Recently, large language models (LLMs), like GPT-4 \cite{ChatGPT4-2023} and LLaMA-3 \cite{LLaMa-3-2024}, have attained noteworthy success in the natural language processing (NLP) domain. This advancement facilitates the development of multimodal LLMs (MLLMs) \cite{MiniGPT-v2, VideoChatGPT-2023} in visual and multimodal domains. Most recent studies \cite{VTimeLLM-2023, TimeChat-2023} demonstrate that training only LoRA (Adapter) \cite{LoRA-2021} can empower MLLMs to seize spans, as depicted in Fig.~\ref{fig:high-level}~(b). However, these MLLM-based methods necessitate intricate multi-stage fine-tuning strategies tailored for VMR. Moreover, they depend on large video datasets annotated with high-quality spans and queries, which are time-consuming and costly to collect.

To alleviate the burden of manual annotations, some previous works \cite{FVLM-2023, wattasserilZeroShotVideoMoment2023, VTG-GPT-2023} propose a zero-shot setting for VMR, which simply reuses off-the-shelf MLLMs trained on other tasks, as shown in Fig.~\ref{fig:high-level}~(c). Nevertheless, these zero-shot methods neglect language biases \cite{linell2004written, DBLP:conf/acl/LiangLZLSM20} in human-annotated queries, including (1) rare words (2) spelling and grammatical errors. As exemplified in \texttt{query2} of Fig.~\ref{fig:high-level}~(a), "\textit{kleenex}" is a rare word compared to "\textit{tissues}", which causes the model \cite{FVLM-2023} to favor common words ("\textit{person}", "\textit{put}", "\textit{box}"), resulting in inaccurate localization. Additionally, grammatical errors ("\textit{Person they}") can misguide the model because only one person appears in the video. Therefore, it is necessary to eliminate the above biases to improve accuracy.

To tackle the above challenges, this paper proposes a zero-shot VMR framework, \textbf{Moment-GPT}, employs frozen MLLMs and without additional fine-tuning on VMR data, as illustrated in Fig.~\ref{fig:high-level}~(d). To mitigate language biases, we utilize LLaMA-3 \cite{LLaMa-3-2024} to optimize the raw query, yielding debiased queries. Given that video contains more redundant information than highly generalized text \cite{GLWHH22}, and inspired by humans linguistically understanding videos \cite{barrett2015saying, rohrbach2013translating}, we apply MiniGPT-v2 \cite{MiniGPT-v2} to obtain frame-level captions from the input video. Then, we compute the similarities between these captions and debiased queries, adaptively producing candidate spans through the proposed span generator. To leverage the video understanding capabilities of MLLMs, and considering that the existing MLLMs \cite{VideoChatGPT-2023, VideoLLaMa-2023} are better at video captioning than VMR, we use Video-ChatGPT \cite{VideoChatGPT-2023} to generate span-level captions. Finally, the span scorer calculates the relevance between span-level captions and debiased queries, followed by post-processing to obtain the final results. To summarize, our main contributions include:

\begin{itemize}
\item 
We propose \textbf{Moment-GPT}, a zero-shot VMR approach using off-the-shelf MLLMs for direct inference.
\item 
We devise a new strategy for query debiasing utilizing LLaMA-3 to enhance performance. In addition, we design the span generator and span scorer to exploit the visual comprehension abilities of MiniGPT-v2 and Video-ChatGPT effectively.
\item 
Extensive experimental results demonstrate that our method outperforms the SOTA MLLM-based and zero-shot approaches on three VMR datasets. Significantly, it also exceeds most supervised models.
\end{itemize}

\section{Related Work}
\label{sec:related_work}

\paragraph{Video moment retrieval.} 
VMR constitutes a promising yet challenging task emphasizing retrieving relevant spans from a video, given a linguistic query. Existing fully-supervised VMR approaches \cite{GPTSee-2023, MH-DETR-2024, MRNet-2024, MomentDETR-2021} conventionally hinge on extensive datasets annotated with queries and corresponding spans for training. However, manually collecting VMR data is costly and labor-intensive; for example, producing QVHighlights \cite{MomentDETR-2021} took around \$17,000 and 1,500 hours. To alleviate reliance on spans, prior studies \cite{CNM-2022, CPL-2022} propose a weakly-supervised setup to learn the unmatched video-query pairs. Further diminishing the dependency on queries, some works \cite{PSVL-2021, gaoLearningVideoMoment2022, PZVMR-2022} introduce unsupervised framework, leveraging k-means clustering or CLIP \cite{CLIP-2021} to generate pseudo queries from videos, or select from a query database. Please note that we align with recent works \cite{Zero-shot-VMR-2023, FVLM-2023} and categorize partially zero-shot methods \cite{PSVL-2021, PZVMR-2022} as unsupervised. This is because these unsupervised approaches heavily rely on videos from VMR-specific datasets, inevitably introducing vision bias from corresponding datasets \cite{liu2024decade}.

\paragraph{Zero-shot video moment retrieval.} To reduce the burden of manual annotation and circumvent vision bias from specific VMR videos, recent works \cite{Zero-shot-VMR-2023, FVLM-2023, VTG-GPT-2023} propose a zero-shot setting repurposing frozen models pretrained on other tasks and without any fine-tuning. Luo et al. \cite{FVLM-2023} generate spans using InternVideo \cite{InternVideo-2022} with refined masks and clustering. Diwan et al. \cite{Zero-shot-VMR-2023} and Wattasseril et al. \cite{wattasserilZeroShotVideoMoment2023} apply a shot-detection technique and CLIP (BLIP-2 \cite{BLIP2-2023}) for span computation, but this strategy is not suitable for scenarios with rapid shot transitions, thereby causing poor overall localization effect. In addition, the above zero-shot methods overlook language bias in the original queries, leading to highly inaccurate predictions on biased queries, as depicted in Fig.~\ref{fig:high-level}~(a). To solve this problem, this paper designs a debiasing strategy using LLM \cite{LLaMa-3-2024} to reduce the bias.

\begin{figure*}[t!]
  \centering
  \includegraphics[width=0.95\linewidth]{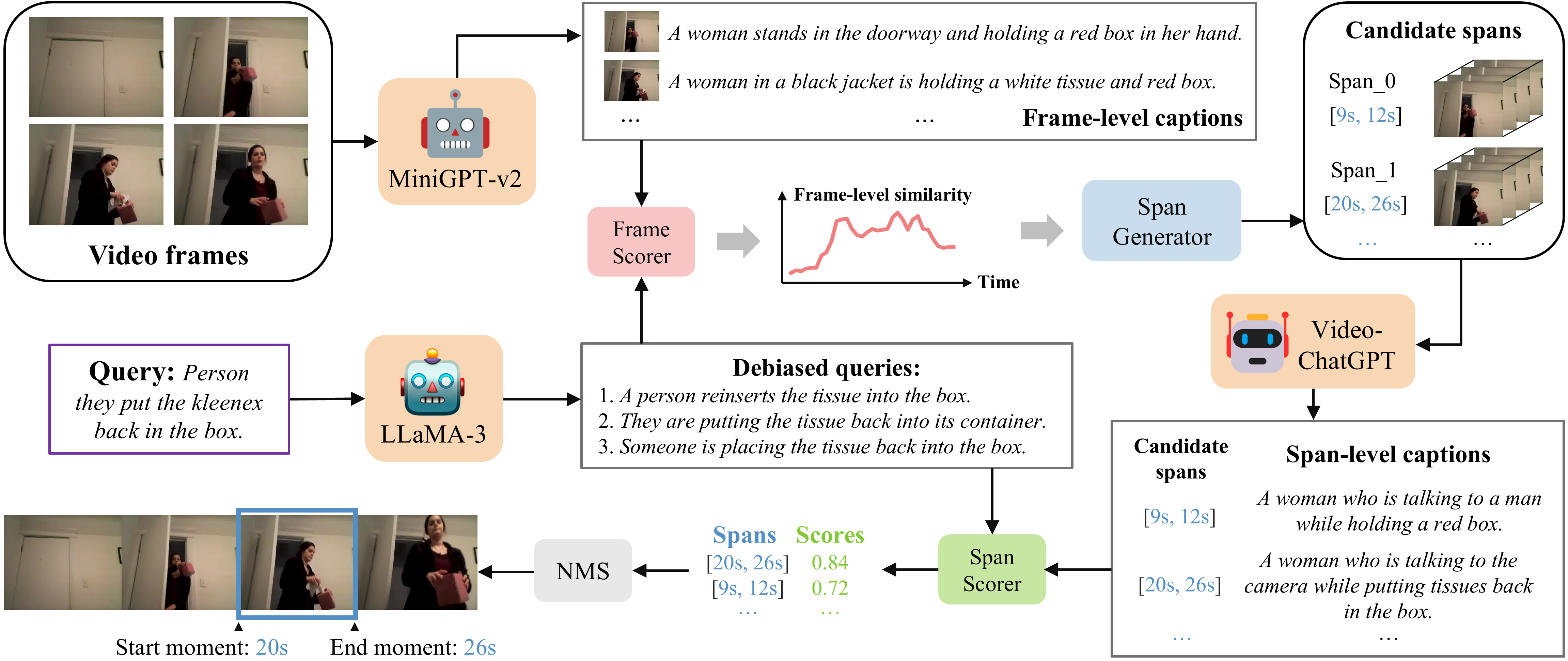}
  \caption{\small
    The overall architecture of Moment-GPT. It first utilizes LLaMA-3 to reduce language bias (Sec.~\ref{subsec:query_debiasing}). Next, construct candidate spans by MiniGPT-v2, frame scorer, and span generator (Sec.~\ref{subsec:generate_candidates}). Finally, select the most relevant spans using Video-ChatGPT, span scorer, and NMS (Sec.~\ref{subsec:select_spans}).
  }
  \label{fig:method}
\end{figure*}

\paragraph{Multimodal large language models.}
Recent LLMs \cite{WizardLM-2023, ChatGPT4-2023, LLaMa-2-2023, LLaMa-3-2024, Hifi-portrait} have attracted widespread attention from researchers due to their remarkable success in various NLP tasks. This success promotes the development of MLLMs in the field of computer vision \cite{MiniGPT4-2023, MiniGPT-v2, Diff-PC-2025}, with representative works like LLaVA \cite{LLaVA-2023} and MiniGPT-v2 \cite{MiniGPT-v2} leveraging appropriate prompts to engage in dialogue, enabling them to summarize images and generate detailed textual descriptions. Subsequently, researchers expand single-frame images into multi-frame videos, with methods like VideoLLMA \cite{VideoLLaMa-2023} and Video-ChatGPT \cite{VideoChatGPT-2023} demonstrating robust video understanding capabilities, yielding superior zero-shot results in tasks including video captioning and video Q\&A. However, these MLLMs \cite{VideoLLaMa-2023, VideoChatGPT-2023} exhibit poor performance in VMR due to the absence of span constraints during training. To address this, recent works \cite{LLaViLo-2023, GroundingGPT-2024} devise complex multi-stage strategies to train LoRA, endowing MLLMs with temporal localization abilities.
Nonetheless, these training strategies are time-consuming and require laborious collection of precise annotations. Moreover, to enhance localization accuracy, some works \cite{VTimeLLM-2023, TimeChat-2023} employ a sliding window to pre-generate overlapping candidate spans, significantly increasing computational costs. Unlike the above methods, this paper leverages frozen MLLMs \cite{LLaMa-3-2024, MiniGPT-v2, VideoChatGPT-2023} to address the above challenges.


\section{Method}

In Sec.~\ref{subsec:overview}, we first define the task and outline our Moment-GPT. Then, we detail the modules and design motivations of Moment-GPT in Sec.~\ref{subsec:query_debiasing}-\ref{subsec:select_spans}.

\subsection{Overview}
\label{subsec:overview}

Given an untrimmed video $V=\{v_i\}_{i=1}^{L_v}$ containing $L_v$ frames, and a textual query $Q=\{q_i\}_{i=1}^{L_q}$ comprising $L_q$ words, the goal of video moment retrieval (VMR) is to predict all temporal spans (segments) $T = \{t^s, t^e\} \in \mathbb{R}^{N_t \times 2}$ semantically relevant to the query: $T = \text{VMR}(V, Q)$, in which $t^s$ and $t^e$ represent the start and end moments, respectively.

Fig.~\ref{fig:method} illustrates the architecture of our proposed Moment-GPT. Concretely, we first apply LLaMA-3 \cite{LLaMa-3-2024} to inspect and rephrase $Q$, thereby mitigating language bias and generating debiased queries $D \in \mathbb{R}^{N_d \times L_d}$. Subsequently, MiniGPT-v2 \cite{MiniGPT-v2} summarizes each image to produce frame-level captions $C^f \in \mathbb{R}^{L_v \times L_f}$. The frame scorer computes frame-wise similarities $S^f \in \mathbb{R}^{N_d \times L_v}$ between $D$ and $C^f$, and then span generator dynamically constructs candidate spans $T^p \in \mathbb{R}^{N_p \times 2}$ from $S^f$. To leverage video understanding abilities of MLLMs, we feed candidates $T^p$ into Video-ChatGPT \cite{VideoChatGPT-2023} to obtain span-level captions $C^s \in \mathbb{R}^{N_p \times L_s}$. After that, the span scorer calculates span-level score $S \in \mathbb{R}^{N_p}$ based on relevance between $C^s$ and $D$, followed by non-maximum suppression (NMS) to derive the most accurate spans $T \in \mathbb{R}^{N_t \times 2}$.

\subsection{Reducing Language Bias}
\label{subsec:query_debiasing}

\begin{figure}[b!]
    \centering
    \includegraphics[width=\linewidth]{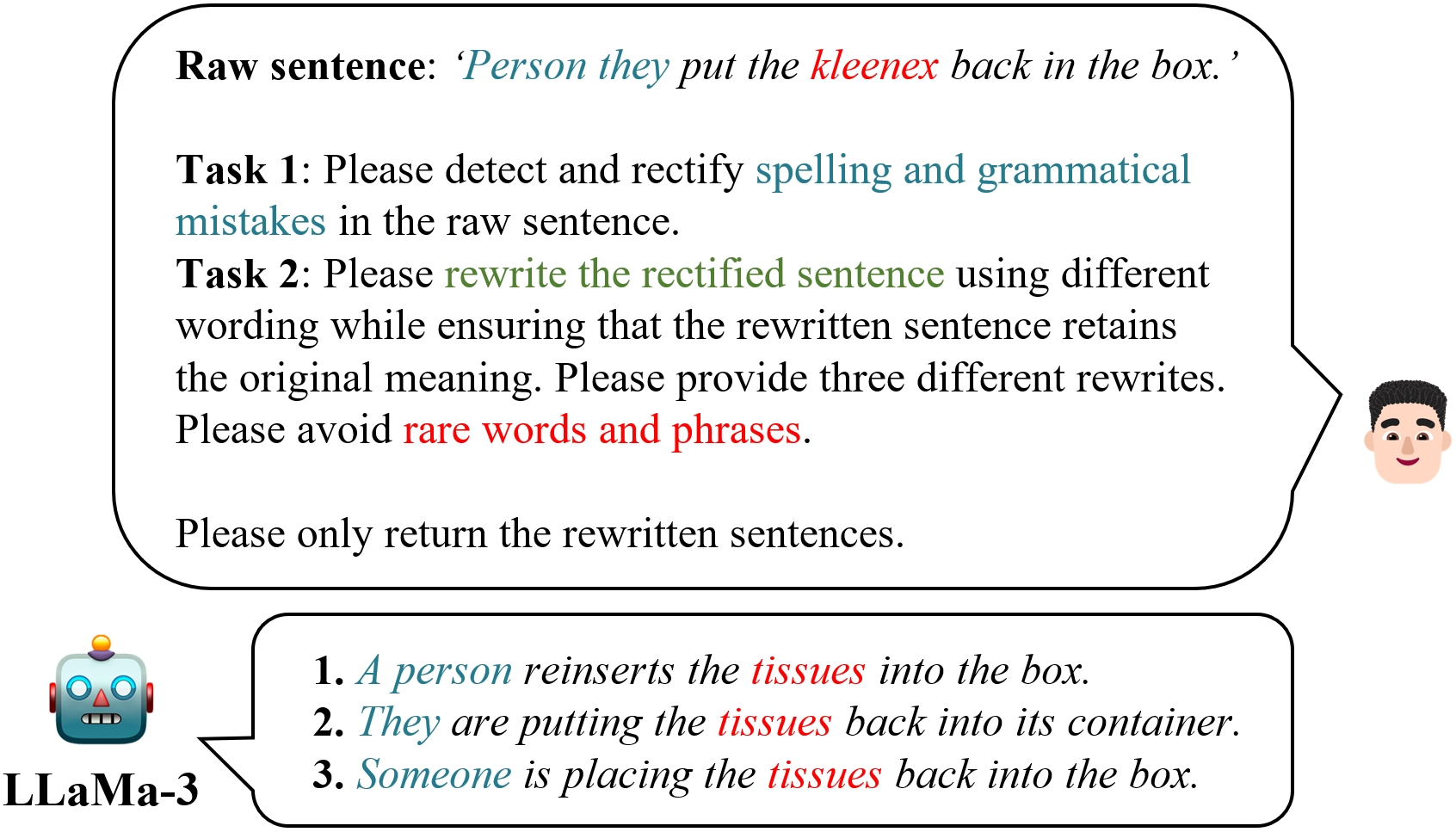}
    \caption{\small
     Reduce language bias in raw query via LLaMA-3. Bold, italics, and colored fonts are utilized only for presentation and are not employed in the code.
    }
    \label{fig:debias}
\end{figure}

Human-annotated queries often contain language bias stemming from the annotators' subjectivity, such as rare words, spelling and grammatical mistakes, which result in the model locating erroneous spans. To address these, we adopt a human-like approach \cite{winograd1972understanding, liddy2001natural} to natural language processing, first using LLaMA-3 \cite{LLaMa-3-2024} to rectify spelling and grammar mistakes in the raw query. Then, synonym substitution \cite{mekala2023echoprompt, kong2024prewrite} is employed to rewrite the rectified query, reducing the occurrence of rare words. However, we observe that very few queries still have rare words after rewriting. Thus, we task LLaMA-3 with emulating diverse expression styles to generate multiple rewritten queries with unchanged semantics, minimizing the likelihood of encountering rare words. Finally, we consolidate the above ideas and craft the following prompt:
\texttt{\small
"Raw sentence: '<Query>' \textbackslash n \textbackslash n 
Task 1: Please detect and rectify spelling and grammatical mistakes in the raw sentence. \textbackslash n Task 2: Please rewrite the rectified sentence using different wording while ensuring that the rewritten sentence retains the original meaning. Please provide three different rewrites. Please avoid rare words and phrases. \textbackslash n  \textbackslash n 
Please only return the rewritten sentences."
}
Here, \texttt{\small<Query>} represents the raw query. We refer to the text returned by LLaMA-3 as debiased queries $D \in \mathbb{R}^{N_d \times L_d}$. Fig.~\ref{fig:debias} depicts the process of query debiasing, where grammatical mistakes ("\textit{Person they}") are rectified to more accurate terms such as "\textit{A person}", "\textit{They}", or "\textit{Someone}". Similarly, the rare word ("\textit{kleenex}") is substituted with a more common alternative ("\textit{tissues}"). Fig.~\ref{fig:visualization}~(top) demonstrates that utilizing these debiased queries can get more precise results.

\begin{figure*}[t!]
  \centering
  \includegraphics[width=0.85\linewidth]{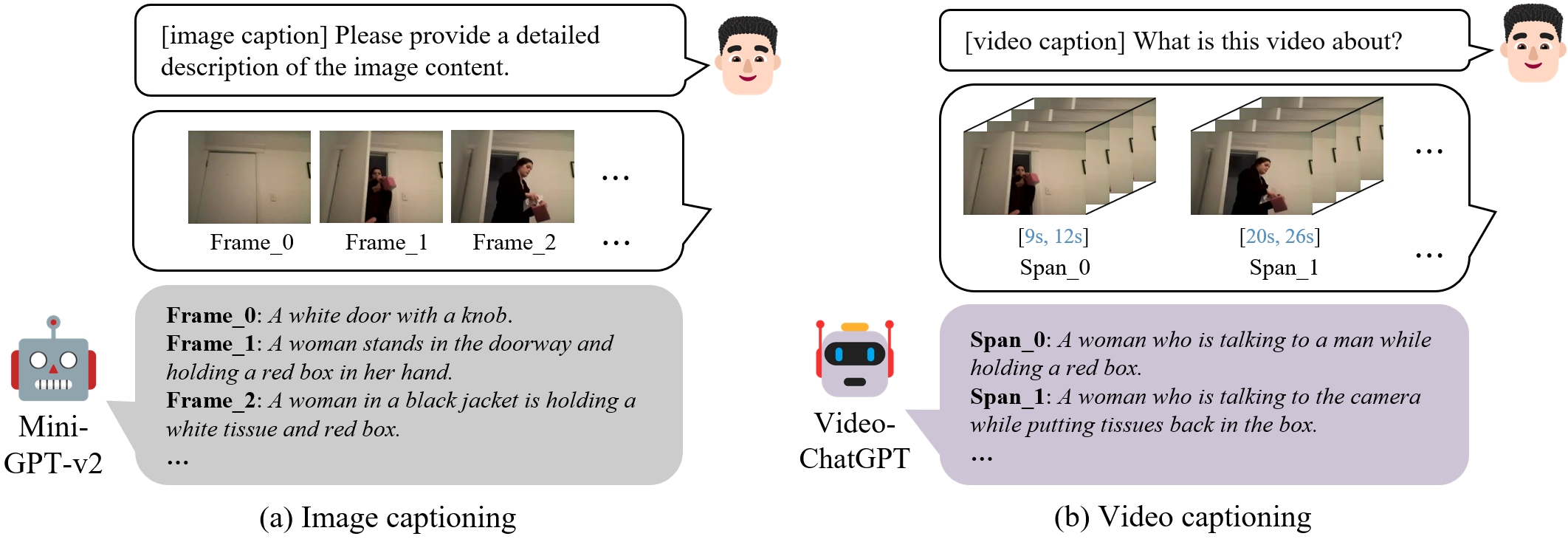}
  \caption{\small
    (a) Image captioning via MiniGPT-v2. (b) Video captioning via Video-ChatGPT. Frame\_N and Span\_N are just for demonstration convenience and do not exist in reality.
  }
  \label{fig:caption}
\end{figure*}

\subsection{Generating Candidate Spans}
\label{subsec:generate_candidates}

To improve localization accuracy, existing methods \cite{VTimeLLM-2023, TimeChat-2023} apply the sliding-window strategy
to construct candidate spans. However, this strategy often leads to the creation of excessively overlapping candidates, resulting in significant computational overhead. In response to this challenge, we propose the following solution (image captioning, frame scorer, and span generator) to adaptively generate candidates $T^p \in \mathbb{R}^{N_p \times 2}$ with low overlap.

\paragraph{Image captioning.}
Drawing from previous zero-shot method \cite{Zero-shot-VMR-2023, wattasserilZeroShotVideoMoment2023}, in our initial experiments, we use BLIP-2 \cite{BLIP2-2023} to calculate $S^f$ between $D$ and each frame in the video. However, this approach is susceptible to visual redundancy \cite{GLWHH22, yu2024she}, resulting in suboptimal outcomes (see Tab.~\ref{tab:5}, row 1). Inspired by human comprehension of visual signals through language \cite{barrett2015saying, rohrbach2013translating}, we utilize MiniGPT-v2 to transform images into highly informative captions $C^f \in \mathbb{R}^{L_v \times L_f}$, as shown in Fig.~\ref{fig:caption}~(a). The instruction as follows: 
\texttt{\small
 "[image caption] Please provide a detailed description of the image content. <Image>". 
}
Here, \texttt{\small[]} is used to emphasize the task within the square brackets, and \texttt{\small<Image>} denotes the actual image. 

\paragraph{Frame scorer.}
Considering that BLIP-2 is a multimodal model, which may not necessarily excel certain language methods in the field of NLP, thus we experiment with various language models \cite{BERT-2019, SimCSE-2021, LLaMa-3-2024}. Eventually, we employ LLaMA-3 to extract pooled features from $D$ and $C^f$, denoted as $X^d =\{X^d_i\}_{i=1}^{N_d} \in \mathbb{R}^{N_d \times d}$ and $X^f =\{X^f_i\}_{i=1}^{L_v} \in \mathbb{R}^{L_v \times d}$, respectively. Next, we compute cosine similarity between $X^d$ and $X^f$ as the frame-level similarities $S^f \in \mathbb{R}^{N_d \times L_v}$:
\begin{equation}
    S^f = \cos(X^d, X^f) = \frac{X^d \cdot X^f}{\|X^d\| \|X^f\|}
\end{equation}
We are surprised to discover that this combination (image captioning and language-based frame scorer) can yield significant gains (see Tab.~\ref{tab:5}, row 5).

\paragraph{Span generator.} 
After obtaining \( S^f \), we design a span generator ($\mathrm{SG}$) to dynamically produce candidates $T^p$. For clarity, we define \( S^{f}_{i} \in \mathbb{R}^{L_v} \) as the cosine score between \( i \)-th debiased feature \( X^d_i \in \mathbb{R}^{d} \) and \( X^f\). Specifically, \( S^{f}_{i,j} \in \mathbb{R}^{1} \) represents the score between \( X^d_i \) and \( j \)-th caption feature \( X^f_j \in \mathbb{R}^{d} \). 

Firstly, we compute the inverse cumulative histogram of \( S^{f}_{i} \), with \(\eta\) bins. We then traverse these bins in reverse order to find the first bin containing at least \(\kappa\) moments, using its left endpoint value as the adaptive threshold \(\gamma\). Next, we iterate through \( S^{f}_{i} \) in temporal order. If \( S^{f}_{i,j} \) exceeds \(\gamma\), the corresponding moment is marked as the starting moment. When the similarities of $\tau$ consecutive moments all fall below $\gamma$, we mark the final moment with a similarity exceeding $\gamma$ as the ending moment. Finally, we repeat the above process to generate a set of candidate spans \( T^p \) from \( S^f \):
\begin{equation}
    T^p = \mathrm{SG}(S^f; \eta, \kappa, \tau)
\end{equation}
where \(\eta\) , \(\kappa\), and \(\tau\) are hyperparameters. For a detailed case analysis, please see \textbf{\textit{appendix}}.

\begin{table*}[t!]
\centering
\scalebox{0.77}{
    \begin{tabular}{ccccccccccc}
    \toprule
    \multirow{2}{*}{Method} & \multirow{2}{*}{MLLM} & \multirow{2}{*}{Setting} & \multicolumn{4}{c}{QVHighlights test} & \multicolumn{4}{c}{QVHighlights val} \\ \cmidrule(l){4-7} \cmidrule(l){8-11} 
     &  &  & R1@0.5 & R1@0.7 & mAP@0.5 & mAP@avg & R1@0.5 & R1@0.7 & mAP@0.5 & mAP@avg \\ \midrule

    VTimeLLM$^\dagger$ \cite{VTimeLLM-2023} & \checkmark  & FS & 47.2 & 29.3 & 47.3 & 27.4 & 48.8 & 29.5 & 49.3 & 26.8 \\
    LLaViLo$^\dagger$ \cite{LLaViLo-2023} & \checkmark & FS & 48.6 & 29.7 & 48.7 & 27.9 & 49.0 & 30.4 & 49.4 & 28.9 \\
    Moment-DETR \cite{MomentDETR-2021} &  & FS & 52.9 & 33.0 & 54.8 & 30.7 & 54.2 & 33.4 & 55.4 & 31.1 \\
    UMT \cite{UMT-2022} &  & FS & 56.4 & 40.8 & 53.1 & 35.4 & - & - & - & - \\ 
    MH-DETR \cite{MH-DETR-2024} &  & FS & 60.1 & 42.5 & 60.8 & 38.4 & 60.9 & 44.9 & 60.8 & 39.3 \\ 
    \midrule
    CNM \cite{CNM-2022} &  & WS & 14.1 & 4.0 & 11.8 & - & - & - & - & - \\
    CPL \cite{CPL-2022} &  & WS & 30.8 & 10.8 & 22.8 & - & - & - & - & - \\
    CPI \cite{CPI-2023} &  & WS & 32.3 & 11.8 & 23.7 & - & - & - & - & - \\ 
    \midrule
    
    PZVMR \cite{PZVMR-2022} &  & US & 14.2 & 4.9 & 15.7 & 4.6 & 12.6 & 5.1 & 16.2 & 5.3 \\
    
    \midrule
    VideoLLaMA \cite{VideoLLaMa-2023} & \checkmark  & ZS & 17.1 & 6.7 & 18.2 & 6.2 & 18.5 & 6.9 & 17.8 & 7.1 \\
    VideoChatGPT \cite{VideoChatGPT-2023} & \checkmark  & ZS & 21.1 & 10.2 & 22.8 & 9.5 & 22.4 & 10.8 & 21.9 & 10.3 \\
    UniVTG \cite{UniVTG-2023} &  & ZS & 25.2 & 9.0 & 27.4 & 10.9 & - & - & - & - \\
    \cite{Zero-shot-VMR-2023} & \checkmark & ZS & - & - & - & - & 48.3 & 31.0 & 47.3 & 28.0 \\
    \cite{wattasserilZeroShotVideoMoment2023}$^\dagger$ & \checkmark & ZS & 52.4 & 31.6 & 51.7 & 29.6 & 53.1 & 32.2 & 52.3 & 30.2 \\
    \textbf{Moment-GPT (Ours)} & \checkmark & ZS & \textbf{58.3} & \textbf{37.7} & \textbf{55.1} & \textbf{35.0} & \textbf{58.9} & \textbf{38.6} & \textbf{55.7} & \textbf{35.9} \\ 
    \bottomrule
    \end{tabular}
}
\caption{\small{
Performance comparison with methods in different settings on QVHighlights. We denote FS for fully supervised, WS for weakly supervised (fine-tuning with both video and query), and US for unsupervised (fine-tuning with VMR-specific video). ZS indicates zero-shot (no VMR data fine-tuning required). "$\dagger$" is our reproduced results.
}}
\label{tab:2}
\end{table*}

\subsection{Choosing Relevant Spans}
\label{subsec:select_spans}
After getting $T^p$, previous methods \cite{VTimeLLM-2023, TimeChat-2023, wattasserilZeroShotVideoMoment2023} use multimodal models \cite{BLIP2-2023, LLaVA-2023} to identify the spans most relevant to the query from $T^p$. However, visual redundancy in the video seriously impacts these methods' accuracy. Moreover, existing MLLMs \cite{VideoChatGPT-2023, VideoLLaMa-2023} have yet to achieve satisfactory results in VMR, while they perform well in video captioning tasks. Therefore, we first employ Video-ChatGPT \cite{VideoChatGPT-2023} for video captioning to leverage the video comprehension capabilities of MLLMs. And then utilize a span scorer and post-processing to select the best matching spans $T \in \mathbb{R}^{N_t \times 2}$.

\paragraph{Video captioning.} 
Fig.~\ref{fig:caption}~(b) outlines the video captioning process, using Video-ChatGPT \cite{VideoChatGPT-2023} to create span-level captions $C^s \in \mathbb{R}^{N_p \times L_s}$. We devise the command: "\texttt{\small [Video caption] What is this video about? <Video>}". Here, \texttt{\small <Video>} represents the video segment corresponding to $T^p$.

\paragraph{Span scorer.}
We apply LLaMA-3 to extract the pooled features of $C^s$, denoted as $X^s \in \mathbb{R}^{N_p \times d}$. Subsequently, we compute the cosine similarity between $X^s$ and $X^d$, resulting in a span-level similarity matrix $S^y \in \mathbb{R}^{N_p \times N_d}$, and then average it to obtain the span-level similarity $S^s \in \mathbb{R}^{N_p}$.
\begin{equation}
    S^s = \mathrm{avg}[ \cos(X^s, X^d) ]
\end{equation}
In our experiments, we observe that in spans containing multiple scenes, the frame-level similarities $S^f$ sometimes exhibits several steep peaks with considerable intervals between them. In these cases, the span generator tends to construct shorter candidates. To alleviate this issue, we calculate the span distance $E^s \in \mathbb{R}^{N_p}$ between the start and end moment and add it to the span-level score $S \in \mathbb{R}^{N_p}$, encouraging the retention of longer spans in the post-processed results.
\begin{equation}
    S = (1 - \lambda) \cdot S^s + \lambda \cdot E^s
\end{equation}
where $\lambda$ represents the distance coefficient.

\paragraph{Post-processing.}
Finally, we sort the candidate spans $T^p$ according to the score $S$ and utilize NMS with an intersection-over-union (IoU) threshold $\sigma$ to eliminate excessively overlapping spans, thereby obtaining the most suitable spans $T \in \mathbb{R}^{N_t \times 2}$:
\begin{equation}
    T = \mathrm{NMS}(T^p, S; \sigma)
\end{equation}

\section{Experiments}
\label{sec:experiments}


\subsection{Datasets and Metrics}
\label{subsec:dataset}
\paragraph{Datasets.} 
To evaluate our proposed method, we conduct experiments on three datasets with different topics: QVHighlights \cite{MomentDETR-2021}, Charades-STA \cite{Charades-STA-dataset-2017}, ActivityNet-Captions \cite{ActivityNet-Captions-dataset-2017}.
\textbf{QVHighlights} encompasses 10,310 queries and 10,148 YouTube videos with diverse topics, including daily vlogs, social news, et al. Moreover, it offers frame-wise saliency scores for video highlight detection (VHD) task to measure the relevance between frames and queries.
\textbf{Charades-STA}, derived from the Charades dataset \cite{Charades-dataset-2016}, comprises 6,670 intricate indoor activity videos and 16,128 query-span pairs. 
\textbf{ActivityNet-Captions} originates from the ActivityNet dataset \cite{ActivityNet-dataset-2015}, encompassing 19,811 outdoor activity videos and 71,957 pairs.
%


\paragraph{Metrics.} 
To ensure fair comparison, we adhere to previous methods \cite{VTimeLLM-2023, FVLM-2023}, employing the following metrics for VMR: R1@n, mAP@m, mAP@avg, and mIoU. Expressly, R1@n signifies the percentage of test queries with at least one correctly localized span (IoU over n) in the top-1 outcomes. Likewise, mAP@m represents the mean average precision for IoU exceeding m, and mAP@avg denotes the average mAP over a set of IoU values [0.5: 0.05: 0.95]. mIoU is mean IoU. Additionally, we adopt HIT@1 and mAP for VHD, where HIT@1 indicates the hit rate for the top-scored moment in the video.

\subsection{Implementation Details}
\label{subsec:implementation}
Following previous works \cite{VTimeLLM-2023, MomentDETR-2021}, we set the frame rates of videos from Charades-STA, ActivityNet-Captions, and QVHighlights to 1, 1, and 0.5, respectively. The employed MLLM models include LLaMA-3-8B, MiniGPT-v2-7B, 
and Video-ChatGPT based on Vicuna-7B-v1.1 \cite{vicuna-2024}. To reduce the randomness of results, we configure the temperatures of LLaMA-3, MiniGPT-v2, and Video-ChatGPT to 0.3, 0.2, and 0.2, respectively. The number of histogram bins $\eta$ is empirically fixed to 10. The hidden dimension $d$ of LLaMA-3 is 4096. We set the number of debiased queries $N_d$ to 3, the counting threshold $\kappa$ to 7, the number of consecutive moments $\tau$ to 5, the distance coefficient $\lambda$ to 0.2, and the IoU threshold $\sigma$ in NMS to 0.9. All experiments are conducted on 1 NVIDIA A100 GPU.

\begin{table*}[t!]
\renewcommand\arraystretch{1.0}
\centering
\scalebox{0.85}{

\begin{tabular}{ccccccccccc}
\toprule
\multirow{2}{*}{Method} & \multirow{2}{*}{MLLM} & \multirow{2}{*}{Setting} & \multicolumn{4}{c}{Charades-STA} & \multicolumn{4}{c}{ActivityNet-Captions} \\ 
\cmidrule(l){4-7} \cmidrule(l){8-11} 

& & & R1@0.3 & R1@0.5 & R1@0.7 & mIoU & R1@0.3 & R1@0.5 & R1@0.7 & mIoU \\ 
\midrule

GroundingGPT \cite{GroundingGPT-2024} & \checkmark & FS & - & 29.6 & 11.9 & - & - & - & - & - \\

VTimeLLM$^\dagger$ \cite{VTimeLLM-2023} & \checkmark  & FS & 55.3 & 34.3 & 14.7 & 34.6 & 44.8 & 29.5 & 14.2 & 31.4 \\

Moment-DETR$^\dagger$ \cite{MomentDETR-2021} &  & FS & 62.1 & 48.2 & 25.3 & 42.3 & 52.6 & 32.5 & 15.3 & 37.8 \\ 
MH-DETR \cite{MH-DETR-2024} &  & FS & - & 51.1 & 26.8 & - & - & - & - & - \\ 

\midrule

CNM \cite{CNM-2022} &  & WS & 50.0 & 36.2 & 14.2 & 34.2 & 51.3 & 30.3 & 11.4 & 33.9 \\
CPL \cite{CPL-2022} &  & WS & 56.0 & 38.1 & 20.3 & 37.8 & 52.4 & 30.9 & 12.0 & 32.6 \\
\cite{huang2023weakly} &  & WS & 59.2 & 44.2 & 22.1 & 39.4 & 54.8 & 32.9 & - & 36.4 \\ 

\midrule
PSVL \cite{PSVL-2021} &  & US & 45.2 & 30.9 & 14.2 & 30.9 & 45.1 & 29.8 & 15.73 & 30.2 \\
\cite{gaoLearningVideoMoment2022} &  & US & 45.3 & 19.8 & 7.9 & - & 45.8 & 25.9 & 12.1 & - \\
\midrule

TimeChat \cite{TimeChat-2023} & \checkmark  & ZS & - & 32.2 & 13.4 & - & - & - & - & - \\
\cite{FVLM-2023}$^\dagger$ & \checkmark & ZS & 53.4 & 36.0 & 19.3 & 34.1 & 45.6 & 27.4 & 12.3 & 28.4 \\
\textbf{Moment-GPT (Ours)} & \checkmark & ZS & \textbf{58.2} & \textbf{38.4} & \textbf{21.6} & \textbf{36.5} & \textbf{48.1} & \textbf{31.1} & \textbf{14.9} & \textbf{30.8} \\ \bottomrule
\end{tabular}
}
\caption{\small{
Comparative evaluation on Charades-STA and ActivityNet-Captions.
}}
\label{tab:1}
\end{table*}

\subsection{Comparison With the State-of-the-Arts} 
\label{subsec:comparison}
To demonstrate the superiority of Moment-GPT, we first compare methods with different settings on QVHighlights (Tab.~\ref{tab:2}). Our approach demonstrates superior performance compared to the current SOTA zero-shot (ZS) method \cite{wattasserilZeroShotVideoMoment2023}, exhibiting at least \textbf{+3.4\%} improvement in each metric. Moreover, it exceeds all FS (fully-supervised), WS (weakly-supervised), US (unsupervised), and MLLM-based methods. Tab.~\ref{tab:1} reports the comparison on Charades-STA and ActivityNet-Captions. Our method achieves remarkable results, outperforming all ZS and US methods. Here, it outperforms the SOTA ZS method \cite{FVLM-2023} by \textbf{+4.8\%} and \textbf{+2.5\%} in R1@0.3 on these two datasets, respectively. Additionally, it achieves competitive performance with FS and WS methods. We attribute the clear benefit of Moment-GPT over the aforementioned approaches to the query debiasing strategy and the effective utilization of MLLMs for generating and selecting spans.

\begin{table}[t!]
    \centering
    \scalebox{0.8}{
    \begin{tabular}{cccc}
      \toprule
      Model & R1@0.5  & R1@0.7  & mIoU \\
      \midrule
      LLaMA-2 \cite{LLaMa-2-2023} & 37.7  & 20.9 & 35.7  \\
      Mistral-7B \cite{Mistral-7B-2023} & 37.2  & 21.1 & 35.4  \\
      LLaMA-3 \cite{LLaMa-3-2024} &\textbf{38.4} & \textbf{21.6} & \textbf{36.5}  \\
      \bottomrule
    \end{tabular}
    }
    \caption{Different LLMs for query debiasing.}
    \label{tab:3}
\end{table}

\begin{table}[t!]
  \centering
  \scalebox{0.8}{
  \begin{tabular}{cccc}
      \toprule
      Model & R1@0.5  & R1@0.7  & mIoU \\
      \midrule
      LLaVA \cite{LLaVA-2023} & 37.1  & 20.3  & 35.2 \\
      MiniGPT-4 \cite{MiniGPT4-2023} & 37.8  & 20.7  & 35.9 \\
      MiniGPT-v2 \cite{MiniGPT-v2} &\textbf{38.4} & \textbf{21.6} & \textbf{36.5}  \\
      \bottomrule
  \end{tabular}
  }
  \caption{Different MLLMs for image captioning.}
  \label{tab:4}
\end{table}

\subsection{Ablation Studies} 
\label{subsec:ablation}
To explore the effectiveness of each module, we conduct extensive ablation studies on Charades-STA.

\paragraph{Impact of LLMs.} 
We first compare different LLMs \cite{LLaMa-2-2023, Mistral-7B-2023, LLaMa-3-2024} for reducing language bias under the same $N_d = 3$ condition (Tab.~\ref{tab:3}). Among them, LLaMa-3 demonstrates the highest performance, attributed to its extensive fine-tuning across diverse tasks and datasets, enabling superior text processing capabilities. 

\paragraph{Image captioning and frame scorer.} 
In Tab.~\ref{tab:4}, we present a comparison of different MLLMs \cite{MiniGPT4-2023, MiniGPT-v2, LLaVA-2023} for image captioning, where MiniGPT-v2 \cite{MiniGPT-v2} attains the best results. Tab.~\ref{tab:5} (left) assesses the influence of the frame scorer. In row 1, BLIP-2 \cite{BLIP2-2023} is directly utilized to compute frame-level similarities between images and debiased queries. Row 2 integrates image captioning and employs BLIP-2's text encoder to obtain similarities. Contrasting rows 1 with 2 illustrates that deriving highly abstracted descriptions via MiniGPT-v2 effectively mitigates visual redundancy, thereby enhancing precision. The comparison between rows 2 and 3 reveals that language model \cite{BERT-2019} is more effective than multimodal model in computing text similarity. Comparing rows 3-5, we finally choose LLaMA-3 for feature extraction in the frame scorer.

\paragraph{Effect of span generator.} 
We scrutinize diverse strategies for span generation, as delineated in Tab.~\ref{tab:6}. For sliding window strategy \cite{TimeChat-2023}, we set the window length to [10, 20, 30] with a sliding step of half the window length. Shot detection adheres to previous works \cite{Zero-shot-VMR-2023, wattasserilZeroShotVideoMoment2023}, leveraging PySceneDetect\footnote{https://www.scenedetect.com/} with consistent settings. The experimental findings show the superior performance of our span generator, demonstrating that our strategy of combining MiniGPT-v2 and frame scorer can effectively generate suitable spans.


\begin{table}[t!]
\centering
\scalebox{0.75}{
    \begin{tabular}{cccc|ccc}
    \toprule
    \multirow{2}{*}{Method} & \multicolumn{3}{c|}{Frame scorer} & \multicolumn{3}{c}{Span scorer} \\ \cmidrule(lr){2-4} \cmidrule(lr){5-7} 
     & R1@0.5 & R1@0.7  & mIoU & R1@0.5 & R1@0.7 & mIoU  \\ \midrule
    BLIP-2 & 34.8  & 19.3  & 32.3 & - & - & -  \\
    BLIP-2-T & 36.9  & 19.7  & 34.2 & - & - & -  \\
    BERT & 37.3  & 20.8  & 35.4 & 37.5 & 20.4 & 35.8 \\ 
    SimCSE & 38.0  & 21.1  & 35.9 & 37.7 & 21.0 & 36.1  \\ 
    LLaMA-3 & \textbf{38.4} & \textbf{21.6} & \textbf{36.5} & \textbf{38.4} & \textbf{21.6} & \textbf{36.5}  \\
    \midrule
    \end{tabular}
}
\caption{\small{
Effect of frame scorer (left) and span scorer (right).
}}
\label{tab:5}
\end{table}

\begin{table}[t!]
  \centering
  \scalebox{0.8}{
  \begin{tabular}{cccc}
      \toprule
      Model & R1@0.5  & R1@0.7  & mIoU \\
      \midrule
      Sliding Window \cite{TimeChat-2023} & 27.4 & 10.3 & 24.7 \\
      Shot Detection \cite{Zero-shot-VMR-2023} & 32.1 & 11.7 & 30.9 \\

      Span Generator & \textbf{38.4} & \textbf{21.6} & \textbf{36.5} \\
      \bottomrule
  \end{tabular}
  }
  \caption{Different span generation methods.}
  \label{tab:6}
\end{table}

\paragraph{Video captioning and span scorer.} 
Tab.~\ref{tab:7} evaluates the efficacy of various MLLMs for video captioning. Row 1 forego video captions and instead utilize the average frame-level similarity as span-level score.
Row 2 extracts frames from video segments and feeds them into image-MLLM (MiniGPT-v2), resulting in poor outcomes. Conversely, row 4 leverages the video understanding capabilities inherent in video-MLLM (Video-ChatGPT \cite{VideoChatGPT-2023}), yielding superior performance. In Tab.~\ref{tab:5} (right), we explore using different models to compute span-level similarity, with LLaMA-3 emerging as the most effective.

\paragraph{Hyperparameter selection.} 
Please see \textbf{\textit{appendix}}.

\begin{table}[t!]
  \centering
  \scalebox{0.8}{
  \begin{tabular}{cccc}
      \toprule
      Model & R1@0.5  & R1@0.7  & mIoU \\
      \midrule
      None  & 30.2 & 12.4 & 28.7 \\
      MiniGPT-v2 \cite{MiniGPT-v2} & 32.4 & 14.3 & 30.7 \\
      VideoLLaMA \cite{VideoLLaMa-2023} & \textbf{38.9} & 21.2 & 36.2 \\
      Video-ChatGPT \cite{VideoChatGPT-2023} & 38.4 & \textbf{21.6} & \textbf{36.5} \\
      \bottomrule
  \end{tabular}
  }
  \caption{Different MLLMs for video captioning.}
  \label{tab:7}
\end{table}


\begin{table}[t!]
    \centering
    \scalebox{0.8}{
    \begin{tabular}{cccc}
      \toprule
      Model & Setting &mAP  & HIT@1 \\
      \midrule
      TimaChat \cite{TimeChat-2023} & FS & 14.5   & 23.9   \\
      Moment-DETR \cite{MomentDETR-2021}  & FS & 35.7 & 55.7 \\
      \midrule
      Moment-GPT & ZS & 35.1  &   60.3  \\
      Moment-GPT$^*$ & ZS & \textbf{36.7}  &   \textbf{62.7}  \\
      \bottomrule
    \end{tabular}
    }
    \caption{Experimental results on the \textit{val} set of QVHighlights for VHD. "*" indicates using the query debiasing strategy.}
    \label{tab:HD}
\end{table}

\subsection{Further Analysis} 
\label{subsec:analysis}

\begin{figure}[t!]
  \centering
  \includegraphics[width=\linewidth]{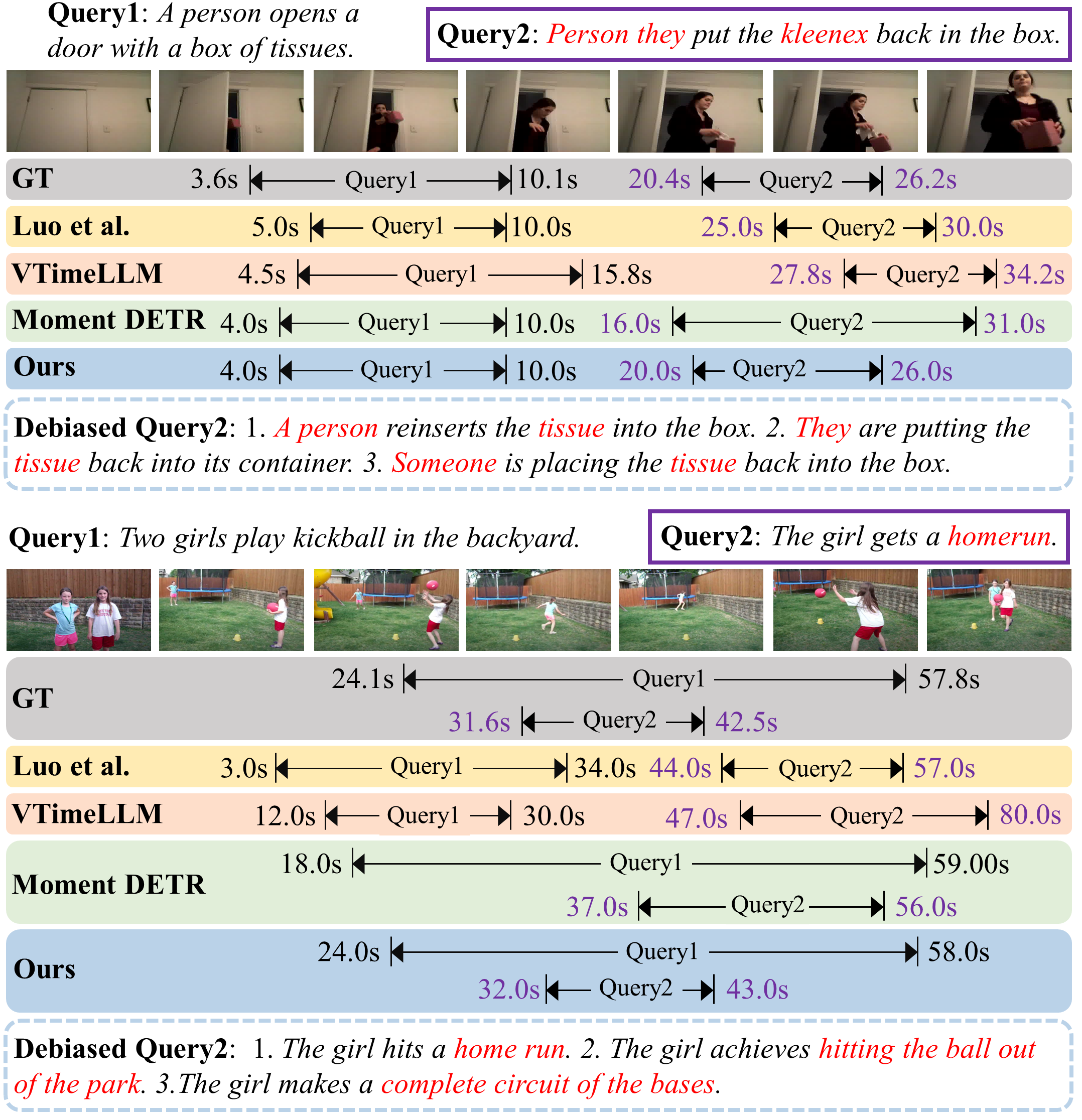}
  \caption{\small
    Qualitative results on Charades-STA (top) and ActivityNet-Captions (bottom). We mark all biased and rewritten words in red.
  }
  \label{fig:visualization}
\end{figure}

\paragraph{Qualitative results.}
To qualitatively verify the effectiveness of our Moment-GPT, we visualize the localization outcomes of ground truth (GT), Luo et al. \cite{FVLM-2023}, VTimeLLM \cite{VTimeLLM-2023}, Moment-DETR \cite{MomentDETR-2021}, and Moment-GPT. Specifically, in Fig.~\ref{fig:visualization}~(top), our model detects the rare word "\textit{kleenex}" in the query and corrects it to "\textit{tissue}". In Fig.~\ref{fig:visualization}~(bottom), the rare word "\textit{homerun}" is interpreted by the model as a more understandable term ("\textit{hitting the ball out of the park}", "\textit{complete circuit of the bases}"). In addition, grammatical errors and spelling errors in the query have been corrected. Moment-GPT achieves more accurate localization than other methods, especially in biased cases. The main reason is that previous methods rely only on the original query, which contains language bias. In contrast, our method could rectify incorrect queries, alleviate the bias derived from annotations, and leverage the video understanding capabilities of MLLMs to achieve more accurate localization. Another case analysis can be found in appendix.

\begin{figure}[t!]
  \centering
  \includegraphics[width=\linewidth]{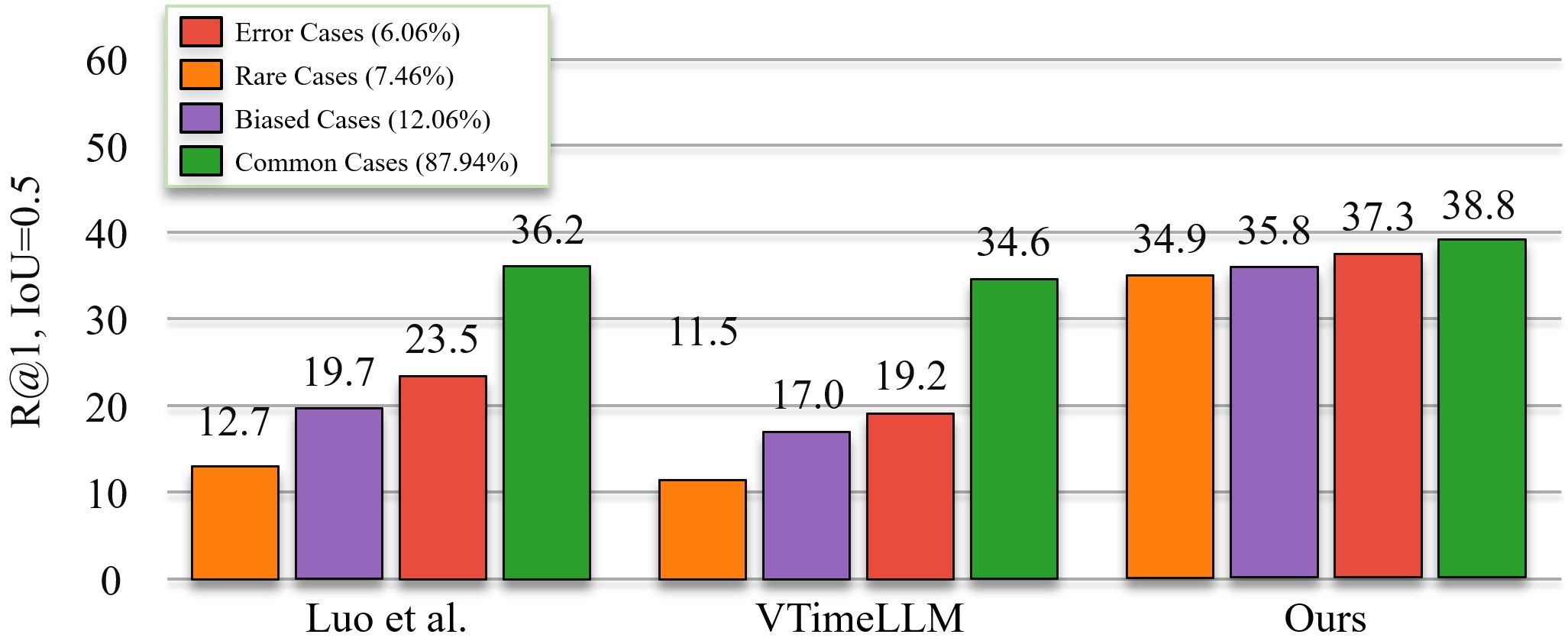}
  \caption{\small
    Analysis of the performance on biased and common cases on Charades-STA.
  }
  \label{fig:biased_case}
\end{figure}

\paragraph{Analysis on query debiasing.}
To further verify the efficacy of query debiasing, we report a comparative evaluation of VHD on QVHighlights. As depicted in Tab.~\ref{tab:HD}, row 3 corresponds to Moment-GPT without debiasing strategy, using the frame-level similarity between the original query and image captions as the saliency score. In contrast, row 4, which applies debiasing, averages the frame-level similarities of debiased queries from the same original query to calculate the saliency score. The experimental results demonstrate that row 4 significantly outperforms row 3, underscoring the effectiveness of this strategy in mitigating language bias. In addition, comparing row 2 with 3, our approach far exceeds the SOTA method \cite{MomentDETR-2021}, which emphasizes the superiority of our model.

\paragraph{Analysis on biased cases.}
In this paper, we categorize queries into different groups: \texttt{\small rare cases}, representing queries with at least one word (noun, verb) occurring less than 10 times; \texttt{\small error cases}, indicating queries containing spelling or grammatical errors; \texttt{\small biased cases}, which encompass the union of rare and error cases; and the remaining queries are termed \texttt{\small common cases}. As illustrated in Fig.\ref{fig:biased_case}, we compare the performance across these four categories on Charades-STA. Our analysis yields the following insights: (1) Previous zero-shot method \cite{FVLM-2023} and MLLM-based method \cite{VTimeLLM-2023} fail to avoid language bias. (2) While VTimeLLM somewhat alleviates language bias through MLLM fine-tuning, its performance remains limited. (3) By optimizing queries using LLM, we successfully reduce language bias, resulting in improved results, particularly on biased cases.


\section{Conclusion}

This paper proposes \textbf{Moment-GPT}, a novel MLLM-based pipeline for zero-shot VMR, which utilizes frozen MLLMs for direct inference, avoiding fine-tuning. Furthermore, it reduces language bias in the original query and effectively leverages MLLMs’ video comprehension abilities. Comprehensive experiments show that Moment-GPT considerably surpasses the SOTA MLLM-based and zero-shot methods on multiple datasets. Additionally, we deliver a comprehensive analysis that elaborates the design choices for each module. Finally, this work can be regarded as a MLLMs-driven multi-agent approach, offering valuable insights for VMR.

\appendix
\section*{Appendix} 

\section{Detailed Case Analysis of Span Generator}

\label{app:generator}
\begin{figure}[h!]
  \centering
  \includegraphics[width=\linewidth]{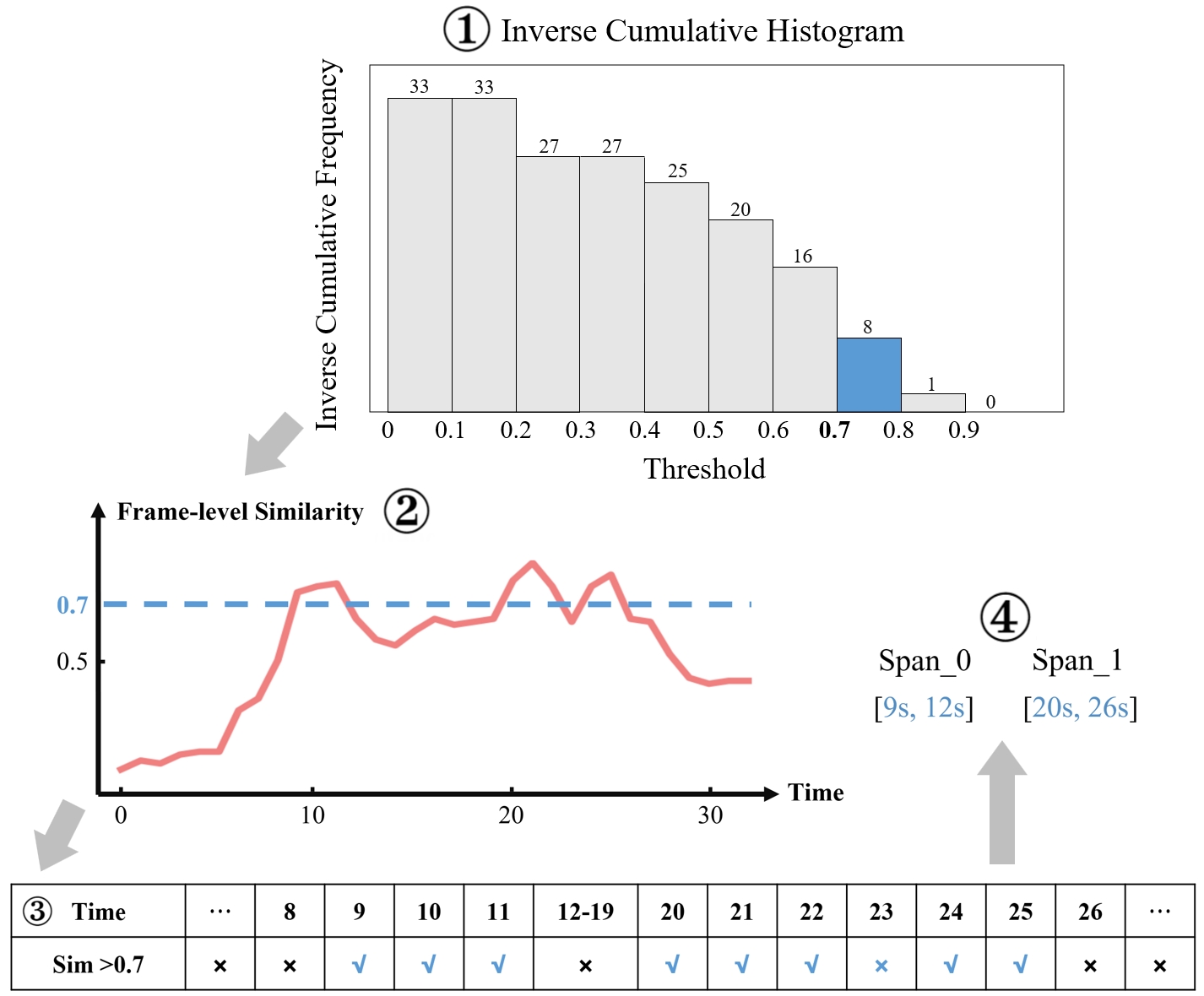}
  \caption{\small
    An example of the span generator.
  }
  \label{fig:generator}
\end{figure}

To better illustrate the span generator, we visualize its workflow (Fig.~\ref{fig:generator}). This example is taken from the Charades-STA \cite{Charades-STA-dataset-2017} dataset; the video ID is "Y1BWP", which contains 33 frames in total. For Charades-STA, we set fps to 1, the number of histogram bins \(\eta\) to 10, the counting threshold \(\kappa\) to 7, and the number of consecutive moments \(\tau\) to 5. We assume that the frame scorer has computed the similarity between frame-level captions and a debiased query, as shown in \ding{173}.

\ding{172} depict that we compute the inverse cumulative histogram of the similarity, with \(\eta = 10\) bins. Next, we find the first bin in reverse order that contains at least \(\kappa = 7\) moments, using its left endpoint value as the adaptive threshold ($\gamma = 0.7$). In \ding{173} and \ding{174}, moment with a similarity above 0.7 are marked with \ding{51}, and those 0.7 or below with \ding{55}. Then, we get spans via the rule: "there will not be $\tau=5$ consecutive \ding{55} in a span". For instance, moments 9 to 11 are all marked as \ding{51}, forming a span. Moments 11 to 20 contains 8 consecutive \ding{55} (more than $\tau$), which cannot form a span. Conversely, there is only one \ding{55} (less than $\tau$) between moments 20 and 25, forming another span. Consequently, we obtain \texttt{\small span\_0:[9s, 12s]} and \texttt{\small span\_1:[20s, 26s]}, as illustrated in \ding{175}. For multiple debiased queries from the same original query, we repeat the above process and finally get a set of candidate spans.

\section{Further Experiments}

\subsection{Efficiency Analysis}
In Tab.~\ref{tab:training cos}, We compare the training cost (TrC), GPU memory usage (GMU), and inference time (InT) on QVHighlights dataset (average video length: 150s) among SOTA fully supervised and zero-shot methods.

As the training-free method, Moment-GPT achieves SOTA VMR performance with 0 training cost. We adhere to the principle of 'make the model larger first and then smaller', so we use 3 MLLMs, which results in slightly higher InT and GPU memory usage. We will soon introduce Tiny-Moment-GPT, which offers higher accuracy, faster speed in zero-shot settings, and lower resource consumption. It will achieve zero-shot VMR using only two small language models (SLMs) \cite{yuan2024tinygptvefficientmultimodallarge,li2023textbooksneediiphi15} cite with tree of thoughts (ToT) for caption/query correction and debiasing.

\begin{table}[t!]
  \centering
  \resizebox{\columnwidth}{!}{ 
  \begin{tabular}{lcccccc}
      \toprule
      \textbf{Setup} & \textbf{Methods} & \textbf{R1@0.5} & \textbf{mAP} & \textbf{TrC} & \textbf{InT (s)} & \textbf{GMU (G)} \\
      \midrule
      ZS & VideoChatGPT \cite{VideoChatGPT-2023} & 22.4 & 10.3 & A100 48h & \textbf{9.8} & \textbf{11} \\
      FS & VTimeLLM \cite{VTimeLLM-2023}     & 48.8 & 26.8 & 4090 40h & 11.2 & 18 \\
      ZS & Wattasseril \cite{li2022blipbootstrappinglanguageimagepretraining}  & 53.1 & 30.2 & \textbf{0}        & 12.7 & 14 \\
      ZS & Ours         & \textbf{58.9} & \textbf{35.9} & \textbf{0} & 16.1 & 32 \\
      \bottomrule
  \end{tabular}
  }
\caption{Comparison of training cost (TrC), GPU memory usage (GMU), and inference time (InT) on the QVHighlights (QVH) dataset.}
  \label{tab:training cos}
\end{table}

\subsection{Out-of-Distribution (OOD) Results}
To further prove Moment-GPT's superiority and generalization, we report out-of-distribution (OOD) comparisons, following DCM ~\cite{yang2021deconfoundedvideomomentretrieval}, on the Charades-STA (C-STA) dataset and the ActivityNet (ANet) dataset. Following VISA ~\cite{li2022compositionaltemporalgroundingstructured}, we report comparisons on the Charades Compositional Dataset (C-CD) and Charades-Caption Generation Dataset (C-CG). Tab.~\ref{tab:ood_comparison} and Tab.~\ref{tab:ood_comparison_singlecol} show that our method achieves SOTA performance on these datasets.

\begin{table}[h!]
  \centering
  \resizebox{\columnwidth}{!}{ 
  \begin{tabular}{ccccccccccc}
      \toprule
      \textbf{Method} & \textbf{Setup} & \multicolumn{2}{c}{\textbf{C-STA OOD-1}} & \multicolumn{2}{c}{\textbf{C-STA OOD-2}} & \multicolumn{2}{c}{\textbf{ANet OOD-1}} & \multicolumn{2}{c}{\textbf{ANet OOD-2}} \\
      \cmidrule(r){3-4} \cmidrule(r){5-6} \cmidrule(r){7-8} \cmidrule(r){9-10}
       &  & R@0.5 & R@0.7 & R@0.5 & R@0.7 & R@0.5 & R@0.7 & R@0.5 & R@0.7 \\
      \midrule
      VSLNeT~\cite{VSLNet-2020} & FS & 17.5 & 8.8 & 10.2 & 4.7 & - & - & - & - \\
      DCM~\cite{yang2021deconfoundedvideomomentretrieval} & FS & 44.4 & 19.7 & 38.5 & 15.4 & 18.2 & 7.9 & 12.9 & 4.8 \\
      Luo~\cite{luo2023zeroshotvideomomentretrieval} & ZS & 40.3 & 18.2 & 38.9 & 17.0 & 18.4 & 6.8 & 18.6 & 7.4 \\
      Ours & ZS & \textbf{45.2} & \textbf{20.6} & \textbf{40.5} & \textbf{18.5} & \textbf{20.0} & \textbf{10.3} & \textbf{16.2} & \textbf{8.7} \\
      \bottomrule
  \end{tabular}
  }
  \caption{OOD Comparisons on Charades-STA (C-STA) and  ActivityNet (ANet)}
  \label{tab:ood_comparison}
\end{table}

\begin{table}[h!]
  \centering
  \resizebox{\columnwidth}{!}{
  \begin{tabular}{cccccccc}
      \toprule
      \textbf{Method} & \textbf{Setup} & \multicolumn{2}{c}{\textbf{C-CD test-ood}} & \multicolumn{2}{c}{\textbf{C-CG novel-composition}} & \multicolumn{2}{c}{\textbf{C-CG novel-word}} \\
      \cmidrule(r){3-4} \cmidrule(r){5-6} \cmidrule(r){7-8}
       &  & R@0.5 & R@0.7 & R@0.5 & R@0.7 & R@0.5 & R@0.7 \\
      \midrule
      VISA \cite{VISA-2022} & FS & - & - & 43.4 & 20.7 & 42.3 & 20.8 \\
      Luo \cite{luo2023zeroshotvideomomentretrieval} & ZS & 39.5 & 17.8 & 40.2 & 16.2 & 45.0 & 21.4 \\
      Ours & ZS & \textbf{46.9} & \textbf{21.2} & \textbf{44.5} & \textbf{21.0} & \textbf{54.8} & \textbf{25.3} \\
      \bottomrule
  \end{tabular}
  }
  \caption{OOD Comparisons on C-CD and C-CG.}
  \label{tab:ood_comparison_singlecol}
\end{table}

\subsection{Hyperparameter Selection}
\paragraph{The number of debiased queries $N_d$.}
We study the influence of varying debiased query numbers $N_d$, as shown in Tab.~\ref{tab:n_d}, where $N_d = 0$ denotes solely utilizing the raw query. The model performs optimally at $N_d = 3$, with a slight decrease observed beyond this value. This suggests that query debiasing can enhance localization accuracy, but excessive rewriting may deviate debiased queries from their original intent, resulting in performance degradation.

\begin{table}[t!]
  \centering
  \scalebox{0.85}{
  \begin{tabular}{cccc}
      \toprule
      $N_d$ & R1@0.5  & R1@0.7  & mIoU \\
      \midrule
      0 & 36.9  & 20.1 & 35.1  \\
      1 & 36.3  & 19.8 & 34.8  \\
      2 & 37.1  & 20.8 & 36.0  \\
      \textbf{3} & \textbf{38.4} & \textbf{21.6} & 36.5  \\
      4 & 38.1  & 21.4 & \textbf{36.6}  \\
      5 & 37.4  & 20.9 & 36.1  \\
      \bottomrule
  \end{tabular}
  }
  \caption{Debiased number $N_d$.}
  \label{tab:n_d}
\end{table}

\begin{table}[t!]
  \centering
  \scalebox{0.85}{
  \begin{tabular}{cccc}
      \toprule
      $\kappa$ & R1@0.5  & R1@0.7  & mIoU \\
      \midrule
      5 & 37.0 & 21.1 & 35.3 \\
      6 & 37.9 & \textbf{21.8} & 35.8 \\ 
      \textbf{7} & \textbf{38.4} & 21.6 & \textbf{36.5} \\
      8 & 37.5 & 21.0 & 35.9 \\
      9 & 37.3 & 20.5 & 35.4 \\
      \bottomrule
  \end{tabular}
  }
  \caption{Counting threshold $\kappa$.}
  \label{tab:kappa}
\end{table}

\begin{table}[t!]
  \centering
  \scalebox{0.85}{
  \begin{tabular}{cccc}
      \toprule
      $\tau$ & R1@0.5  & R1@0.7  & mIoU \\
      \midrule
      4 & 37.8 & 20.1 & 35.9 \\ 
      \textbf{5} & \textbf{38.4} & \textbf{21.6} & \textbf{36.5} \\
      6 & 37.3 & 21.4 & 36.1 \\
      8 & 36.8 & 19.7 & 35.4 \\
      10 & 35.1 & 19.0 & 33.6 \\
      \bottomrule
  \end{tabular}
  }
  \caption{Consecutive moments $\tau$.}
  \label{tab:tau}
\end{table}

\begin{table}[t!]
  \centering
  \scalebox{0.85}{
  \begin{tabular}{cccc}
      \toprule
      $\lambda$ & R1@0.5  & R1@0.7  & mIoU \\
      \midrule
      0 & 37.9 & 21.0 & 36.1 \\
      0.1 & 38.1 & 21.5 & \textbf{36.7} \\
      \textbf{0.2} & \textbf{38.4} & \textbf{21.6} & 36.5 \\
      0.3 & 37.8 & 21.3 & 36.3 \\
      0.4 & 37.4 & 20.7 & 35.8 \\
      \bottomrule
  \end{tabular}
  }
  \caption{Distance coefficient $\lambda$.}
  \label{tab:lambda}
\end{table}

\begin{figure*}[t!]
  \centering
  \includegraphics[width=0.95\linewidth]{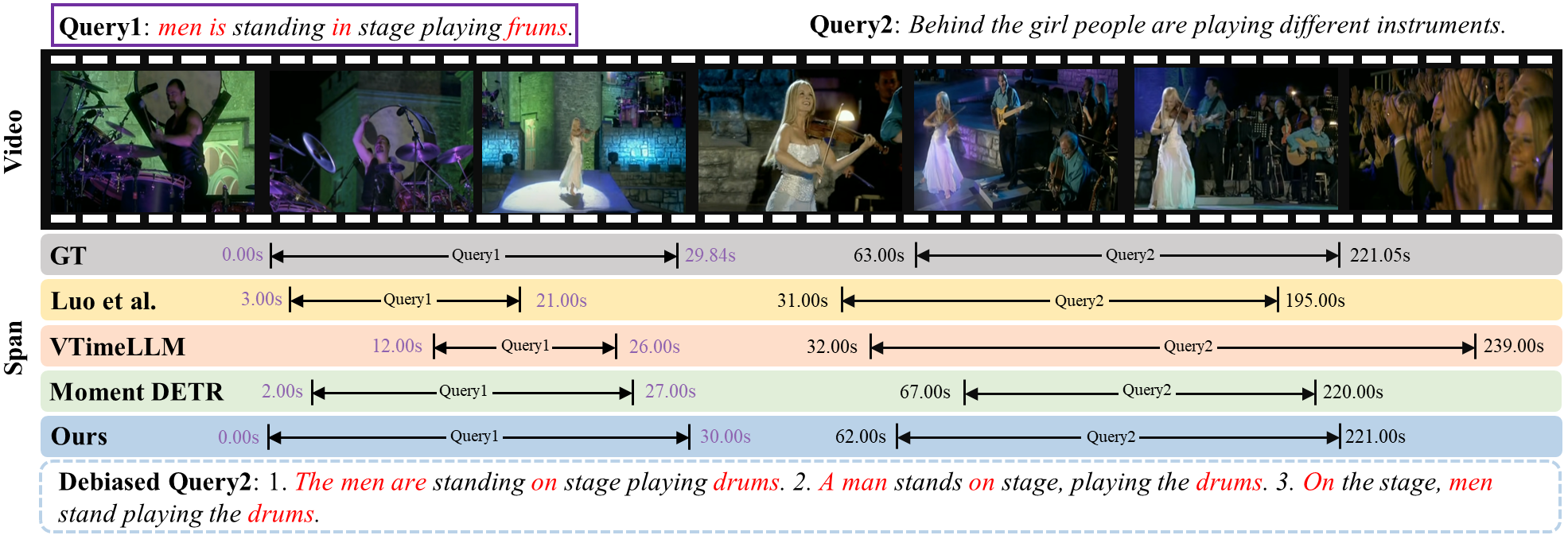}
  \caption{\small
    An example of a query with spelling and grammatical errors. We red out all biased and debiased words.
  }
  \label{fig:vis}
\end{figure*}

\paragraph{Hyperparameter in span generator.}
In Tab.~\ref{tab:kappa} and \ref{tab:tau}, we investigate the impact of two pivotal hyper-parameters on the span generator: the counting threshold $\kappa$ and the number of consecutive moments $\tau$, yielding optimal efficacy at 7 and 5, correspondingly.

\paragraph{Distance coefficient $\lambda$.}
Tab.~\ref{tab:lambda} depicts the effect of the distance coefficient $\lambda$. Compared to the scorer that disregards distance ($\lambda=0$), setting $\lambda=0.2$ encourages the retention of longer spans during post-processing, thereby enhancing localization accuracy.

\paragraph{IoU threshold $\sigma$.} 
Since NMS does not affect the value of R1@n, we report the impact of the IoU threshold $\sigma$ on the mIoU metric in Tab.~\ref{tab:iou}. Compared to without NMS ($\sigma = 1.0$), using $\sigma = 0.9$ can yield certain gains, indicating that NMS effectively reduces overlapping spans.

\subsection{Oracle Bounds}
Following previous works \cite{Zero-shot-VMR-2023, wattasserilZeroShotVideoMoment2023}, in Tab.~\ref{tab:bound}, we compute the oracle bounds to gauge the effectiveness of video captioning and the span scorer. The oracle bounds represent a set of pre-set non-overlapping candidate spans containing the ground truth. Experimental results demonstrate a substantial enhancement in predictive accuracy when using the oracle bound.

\begin{table}[h!]
  \centering
  \scalebox{0.85}{
  \begin{tabular}{cc}
      \toprule
      $\sigma$ & mIoU \\
      \midrule
      0.8 & 36.2  \\
      0.85 & 36.4  \\
      0.9 & \textbf{36.5}  \\
      0.95 & 36.3  \\
      1.0 & 35.9  \\
      \bottomrule
  \end{tabular}
  }
  \caption{IoU threshold $\sigma$.}
  \label{tab:iou}
\end{table}

\subsection{Visualization}
\label{app:vis}
Fig.~\ref{fig:vis} shows an example of an original query with spelling and grammatical errors. In \texttt{\small query1}, "\textit{frums}" is a misspelled word, and "\textit{men is}" and "\textit{standing in stage}" contain innocuous grammatical errors. Zero-shot method (Luo et al. \cite{FVLM-2023}) and MLLM-based method (VTimeLLM \cite{VTimeLLM-2023}) are affected by language bias, and both have serious localization errors. In addition, we found that VTimeLLM has severe hallucinations. The video length of this example is only 230s, but it predicts that the end moment of \texttt{\small query2} is 239s. In comparison, our proposed Moment-GPT rewrites the incorrect word in the query as "\textit{drums}" and corrects the grammatical error, making the localization more accurate.

\begin{table}[t!]
    \centering
    \scalebox{0.85}{
        \begin{tabular}{@{}cccc@{}}
        \toprule
        Method & R1@0.5 & R1@0.7 & mIoU \\ 
        \midrule
        Zero-shot & 38.4 & 21.6 & 36.5 \\
        Oracle Bound & 68.5 & 47.2 & 67.8 \\ 
        \bottomrule
        \end{tabular}
    }
    \caption{Results using oracle bounds on Charades-STA. Oracle bounds mean a preset set of non-overlapping candidate spans that contain ground truth.
    }
    \label{tab:bound}
\end{table}

\section{Discussion}
\label{app:discussion}

\subsection{Limitations and Future Works}
\label{app:limitation_future}
Due to computational limitations and the current performance constraints of MLLMs, we adopt a compromise strategy for zero-shot VMR. That is obtaining highly abstracted captions and rephrased queries from MLLMs, followed by similarity computation. Nevertheless, a more efficacious approach involves employing a single MLLM to directly extract features from text and video simultaneously, ensuring that the two modalities share a more aligned semantic space. This approach circumvents the conversion to human-readable text before computing similarity, thereby minimizing encoding biases \cite{xiao2024verbalized} and information loss. However, there are currently no open-source MLLMs suitable for this task. Future work could leverage ultra-large MLLMs (e.g., GPT-4o \cite{ChatGPT-4o}), an early version of AGI (artificial general intelligence), to address the above challenges.

\subsection{Broader Impacts}
\label{app:impact}
This work contributes to zero-shot VMR, with potentially positive or negative implications depending on the application context. On the one hand, such models can find utility in video surveillance and thus may lead to problematic usage. On the other hand, our method solely leverages frozen MLLMs, obviating the need for costly datasets and time-consuming training, thus reducing carbon emissions. Additionally, this work can be regarded as a MLLMs-driven multi-agent approach, offering valuable insights for future AGI research.

\subsection{Ethical Considerations}
\label{app:ethics}
Our approach follow \cite{Saravia_Prompt_Engineering_Guide_2022} for prompt design and use MLLMs \cite{LLaMa-3-2024, MiniGPT-v2, VideoChatGPT-2023} that have passed the poisoning test. In our experiments, we did not observe MLLMs generating harmful or offensive content, nor did we observe any jailbreaking responses from MLLMs.

\clearpage

\bibliography{aaai25}

\end{document}